\documentclass[a4paper,11pt]{article}
\usepackage{pos}

\title{An updated survey of Active Galaxies with the HAWC gamma-ray observatory }
 \ShortTitle{An updated survey of Active Galaxies with the HAWC gamma-ray observatory }

\author*[a]{Fernando Ureña-Mena}
\author[a]{Alberto Carramiñana}
\author[a]{Daniel Rosa-González}
\author[b]{Anna Lia Longinotti}

\affiliation[a]{Instituto Nacional de Astrofísica, Óptica y Electrónica (INAOE)\\
  Luis Enrique Erro 1, Tonantzintla, Puebla, Mexico}

\affiliation[b]{Instituto de Astronomía, Universidad Nacional Autónoma de México (IA-UNAM),\\ Ciudad de México, Mexico\\
}

\onbehalf{for the HAWC collaboration} 

\emailAdd{furena@inaoep.mx}
\abstract{We present an update of the survey of Active Galaxies with the High Altitude Water Cherenkov (HAWC) gamma-ray observatory. This work adds 567 days of HAWC data to the previously published survey, providing a refined analysis of an updated total exposure of 2090 days. The sample includes 138 nearby AGNs from the 3FHL catalog. We fit a modified power-law to their very high energy spectra, including the exponential attenuation caused by the Extragalactic Background Light. We found four sources with significant detections (above 5$\sigma$): the radio galaxy M87 and the BL Lac objects Mkn 421, Mkn 501 and 1ES 1215+303. We also report eight sources with a marginal detection (between 3$\sigma$ and 5$\sigma$) of which seven are classified as BL Lac objects and one as a radio galaxy. }

\ConferenceLogo{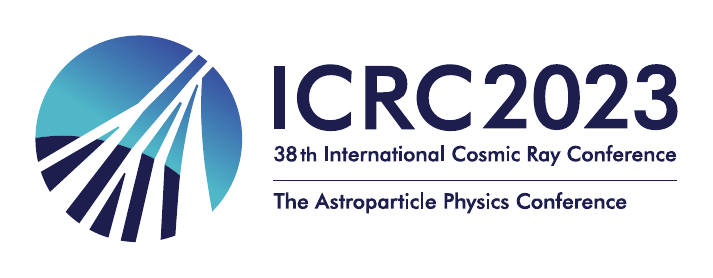}

\FullConference{%
38th International Cosmic Ray Conference (ICRC2023)\\
  26 July - 3 August, 2023\\
  Nagoya, Japan}

\begin{document}
\maketitle

\section{Introduction}

Active Galactic Nuclei (AGN) are luminous extragalactic sources powered by accreting supermassive black holes hosted in the center of  galaxies \cite{netzer2013}. Some of them present collimated emission in form of relativistic jets \cite{blandford2019}, which may extend up to Mpc scales reaching the intergalactic medium out of their host galaxies. Two subclases of jetted AGNs, blazars and radio galaxies, constitute the most common extragalactic gamma-ray sources \cite{4fgl2020}. This gamma-ray emission is produced by non-thermal processes in the relativistic jet and its apparent luminosity is enhanced by Doppler boosting \cite{dermer2009}. \\

Gamma-ray emission from extragalactic sources is heavily attenuated by photon-photon interaction with the Extragalactic Background Light (EBL)\cite{dwek2013,albert2022}, which is a diffuse emission constituted by all the electromagnetic radiation emitted by galaxies in the history of the Universe. This attenuation increases for the most distant sources, as well as for the the highest energies. \\

The High Altitude Water Cherenkov (HAWC) is a gamma-ray observatory, which has been in almost continuous operations ($>95\%$ duty cycle) since 2014. It is located at an altitude of 4100 m in the state of Puebla, Mexico, in the area of the Volcán Sierra Negra mountain. The array consists of 300 Water Cherenkov Detectors (WCD) and is able to study gamma-rays in a wide energy range ($\sim 0.1-100$ TeV).  However, due to the EBL attenuation, most of AGNs can only be studied at the lowest TeV energies. \\

In a recently published survey of AGNs \cite{albert2021}, 1523 days of HAWC data were analyzed looking for evidence of gamma-ray emission from a sample of 138 nearby ($z<0.3$) active galaxies, which were selected from the Third Catalog of Hard Fermi-LAT sources (3FHL) \cite{3fhl2017}. In that work, we found five sources with a significance above $3\sigma$, the BL Lac objects Mrk 421 ($65\sigma$), Mrk 501 ($17 \sigma$), 1ES 1215+303 ($3.6 \sigma$) and VER J0521+211 ($3.2 \sigma$), in addition to the radio galaxy M87 ($3.6 \sigma$).\\

\section{Data and Methodology}

We considered the exact same sample of 138 nearby ($z<0.3$) AGNs  as in the previous HAWC survey of active galaxies \cite{albert2021}. However, the current data set consists of 2090 days covering from 2014 November 26 to 2021 January 14. This represents an increase of $37.2\%$ in time with respect to the previous study.\\

As we did in the previous work, we fit the following spectral model to the whole AGN sample:

\begin{equation}
\label{eq:spc}
\left(\frac{\mathrm{d}N}{\mathrm{d}E}\right)_{obs}=K \left(\frac{E}{1 \ \mathrm{TeV}}\right)^{-\alpha}e^{-\tau(E,z)},
\end{equation}

fixing the spectral index $\alpha=2.5$  and fitting only the normalization ($K$). The  exponential term accounts for the attenuation produced by the photon-photon interactions with the EBL, which follows the model by \cite{dominguez2011}. \\

Then, we calculated the test statistic ($TS$) for each source, which is defined in terms of the log-likelihood ratio between the best fit point source  ($\mathcal{L}_1$, source+background model) and the null hypothesis ($\mathcal{L}_0$, background-only model),

\begin{equation}
    TS=2 \ln \left( \frac{\mathcal{L}_1}{\mathcal{L}_0} \right).
\end{equation}

For those sources with $TS>9$, we performed a second analysis fitting both the normalization $K$ and the spectral index $\alpha$ as free parameters.

\section{Results and discussion}

\begin{figure}
        \centering
        \includegraphics[width=\textwidth]{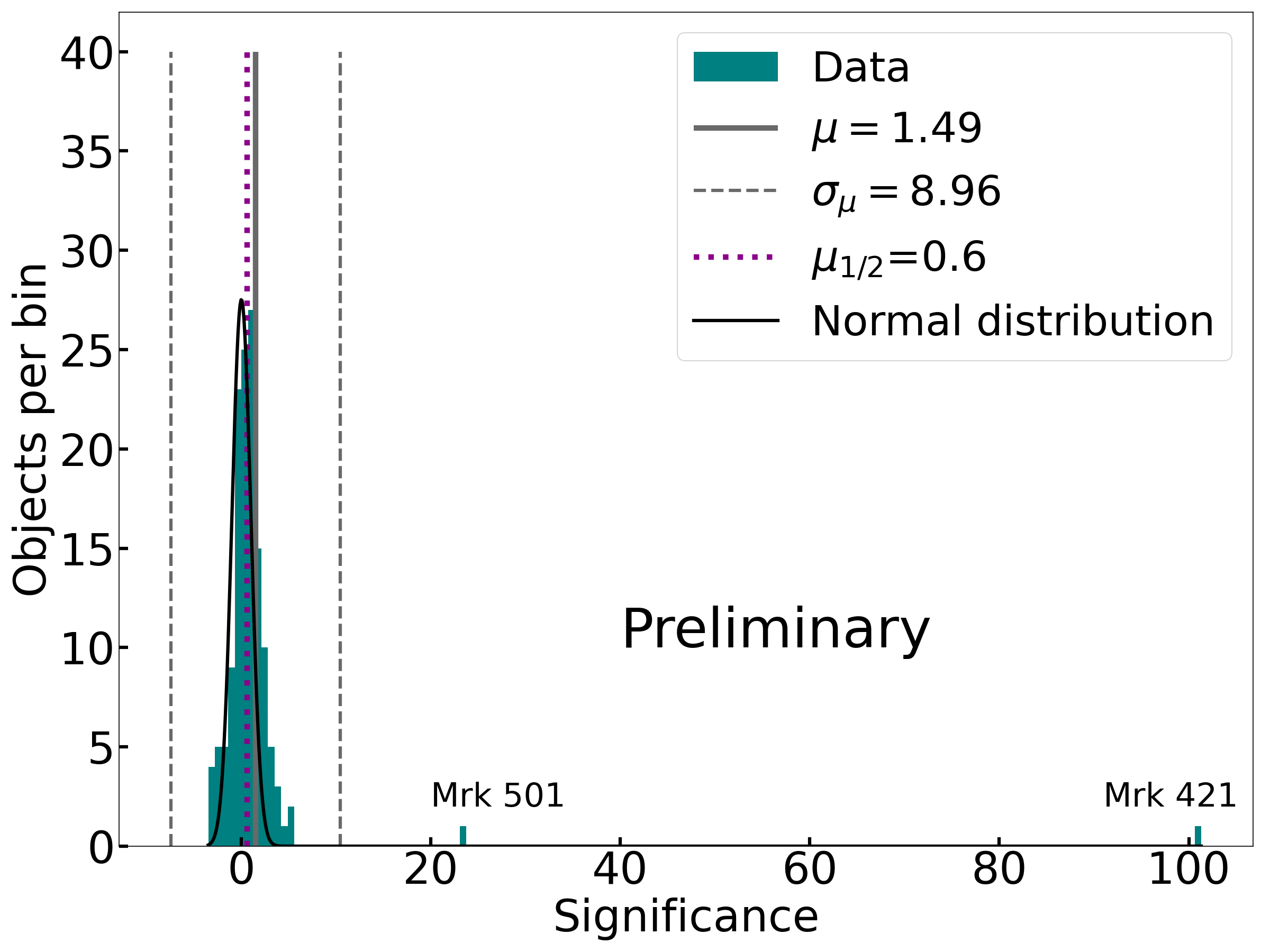}
        \caption{Histogram of significances  (defined as $s=\sqrt{TS}$) for the $\alpha=2.5$ study, including the two most significant sources (Mrk 421 and Mrk 501). }
        \label{fig:hist_with}
    \end{figure}

\begin{figure}
        \centering
        \includegraphics[width=\textwidth]{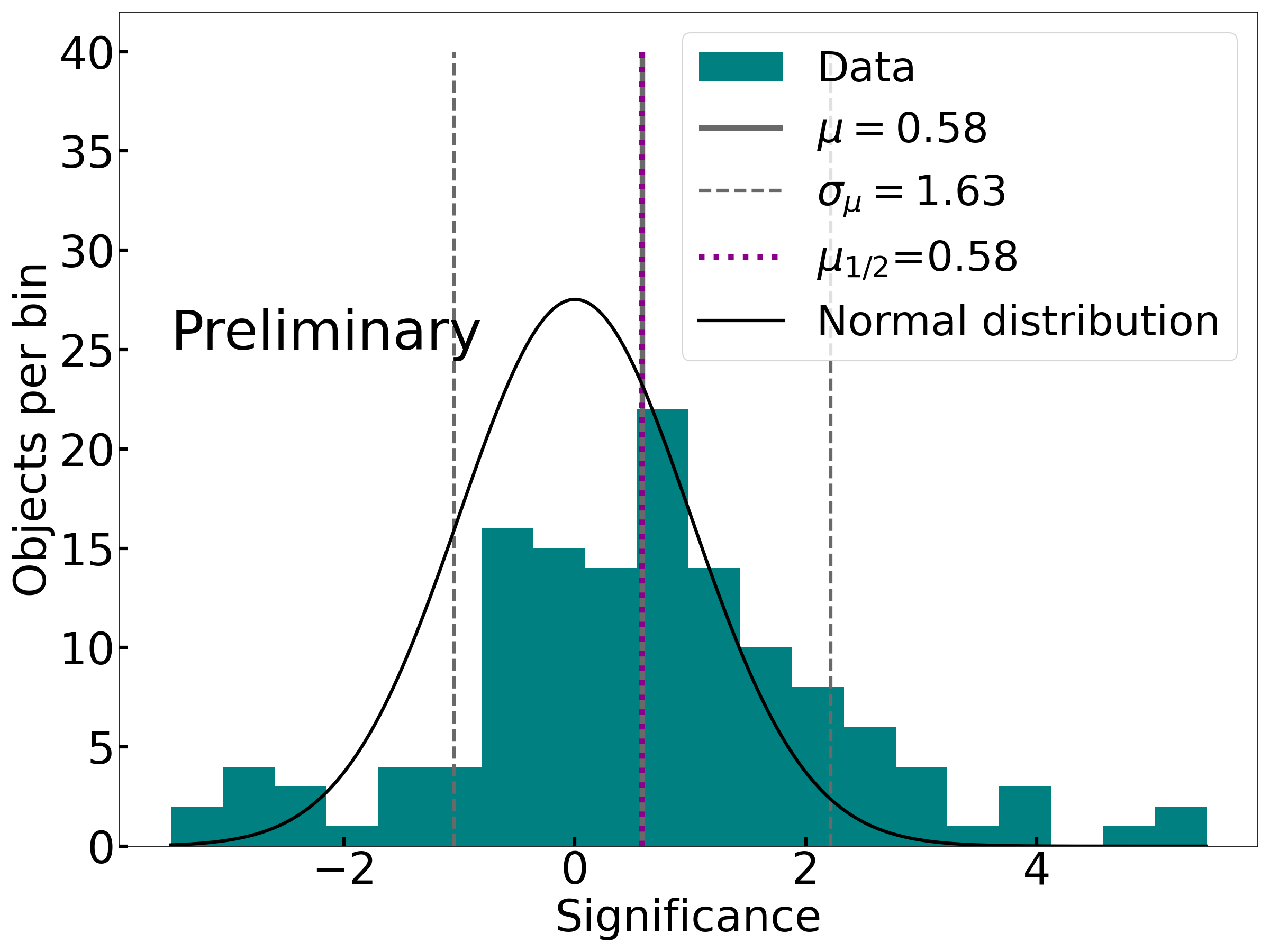}
        \caption{Histogram of significanes  (defined as $s=\sqrt{TS}$) for the $\alpha=2.5$ study, excluding the two most significant sources (Mrk 421 and Mrk 501). }
        \label{fig:hist_without}
    \end{figure}

After carrying out the initial analysis with the spectral index fixed to 2.5, we found 12 sources with $TS>9$. From these sources, ten are classified as BL Lac objects (a subclass of blazars) and two as radio galaxies. Moreover, four presented a $TS>25$. This represents a significant improvement in the number of detections compared to previous results, in which we found only five sources with $TS>9$. The 12 HAWC sources are listed in Table \ref{tab:fixed} along with their TS values. In addition, Figure \ref{fig:hist_with} shows the histogram of  significances (defined as $s=\sqrt{TS}$) for the whole sample. In order to have a better insight, Figure \ref{fig:hist_without} depicts the same histogram but excluding the two most significant sources, Mrk 421 and Mrk 501.\\

\begin{table}
    \centering
    \begin{tabular}{c c c c}
    \hline
        Detections ($>5 \sigma$) & $TS$ & Marginal detections $(3\sigma-5 \sigma)$ & $TS$\\ \hline
         \textbf{BL Lac objects}: & & \textbf{BL Lac objects}: & \\
         Mrk 421  & $10265.12$ & ZS 0214+083 & $14.28$ \\
         Mrk 501 &$ 558.08$ & PKS 0422+00 & $9.90$ \\
         1ES 1215+303 & $26.28$& VER J0521+211 & $16.32$ \\
         \textbf{Radio Galaxies}:& &  RX J0648.7+1516 & $11.78$  \\
         M87 & $ 29.87$ & RX J1100.3+4019 & $ 13.54$ \\
         & & PG 1218+304 & $ 23.72$\\
         & & W Comae & $ 10.08$\\
         & & \textbf{Radio Galaxies}: & \\
         & & 3C 264  & $ 9.56$\\ \hline

    \end{tabular}
    \caption{AGNs in our sample with $TS>9$ in the $\alpha=2.5$ study. }
    \label{tab:fixed}
\end{table}
    
\begin{figure}
    \centering
    \includegraphics[width=\textwidth]{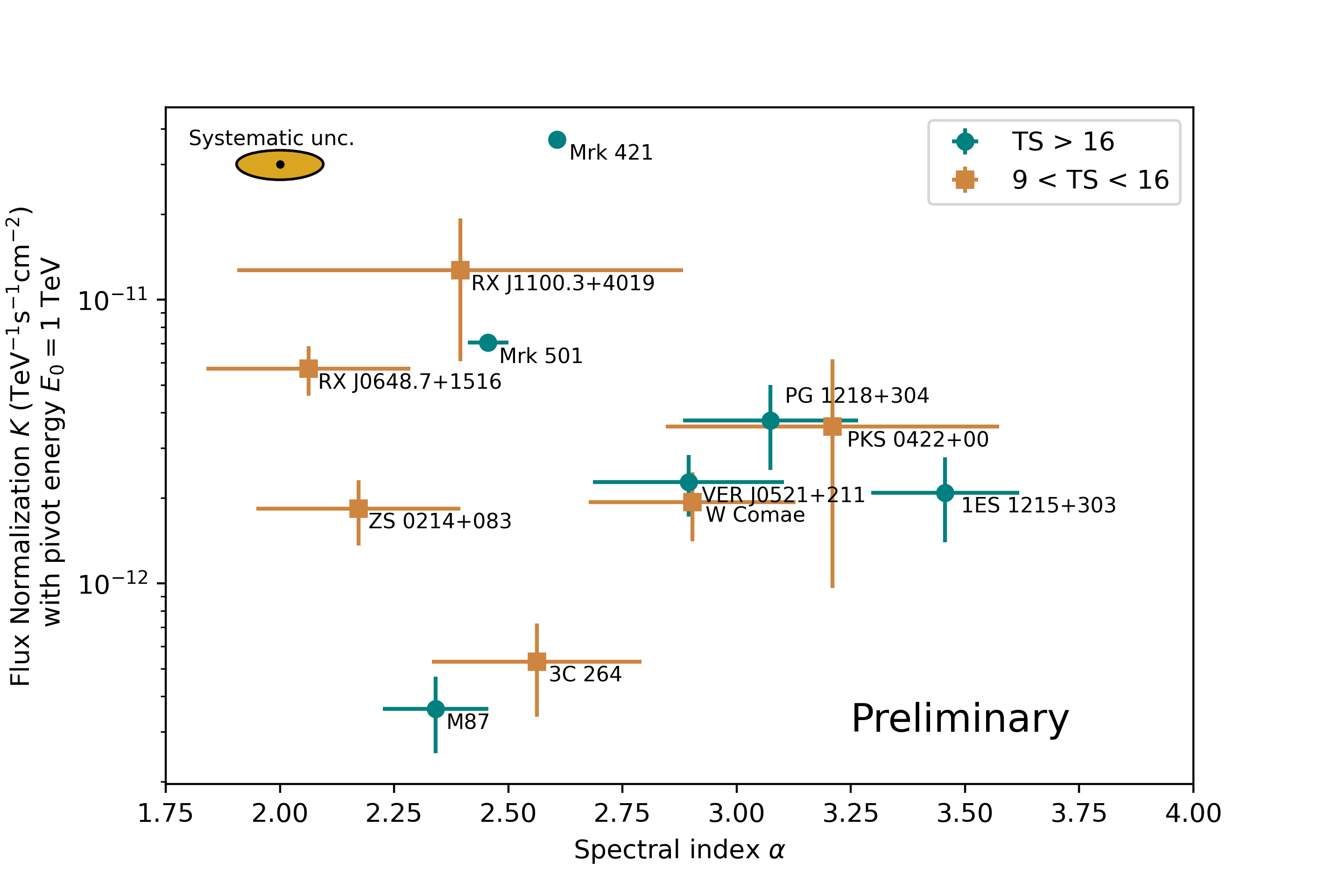}
    \caption{Best fit values of the spectral parameters ($K$,$\alpha$) for the sources in our sample with $TS>9$. The error bars are purely statistical, but the ellipse represents the systematic uncertainties computed for Mrk 421.
 }
    \label{fig:ind_K}
\end{figure}

As we mentioned in the previous section, the 12 sources with $TS>9$ were then fitted by allowing both the normalization and the spectral index to vary, as defined in Eq. \ref{eq:spc}. We obtained an increase in $TS$ for every source in the sub-sample,  as can be seen in Table \ref{tab:free}. It is worth mentioning that the search on the updated data set led to the detection of a new source with $TS>25$, the blazar PG 1218+304. However, this source is just 0.88 degrees from another source in our sample, 1ES 1215+303, which implies that we can not rule out a possible contamination between these two objects.  Finally, Figure \ref{fig:ind_K} shows the best fit values for the gamma spectral parameters in the most significant sources in our sample.

\begin{table}
    \centering
    \begin{tabular}{c c c c}
    \hline
        Detections ($>5 \sigma$) & $TS$ & Marginal detections $(3\sigma-5 \sigma)$ & $TS$\\ \hline
         Mrk 421  & $10332.3$ & ZS 0214+083 & $15.474$ \\
         Mrk 501 &$ 558.592$ & PKS 0422+00 & $11.541$ \\
         1ES 1215+303 & $43.8617$& TXS 0518+211 & $18.2545$ \\
          M87 & $ 30.8741$ &  RX J0648.7+1516 & $13.4317$  \\
         PG 1218+304 & $ 28.4891$ & RX J1100.3+4019 & $ 13.5593$ \\
         
         & & W Comae & $ 11.7698$\\
         & & 3C 264  & $ 9.59461$\\ \hline

    \end{tabular}
    \caption{Sources in our sample with $TS>9$, with the $TS$ values obtained after fitting both the normalization $K$ and spectral index $\alpha$.}
    \label{tab:free}
\end{table}

\section{Concluding remarks}

After analyzing 2090 of HAWC data we found 12 active galaxies, from a sample of 138 selected from the Fermi FHL catalogue, with $TS>9$.  After obtaining  intrinsic gamma spectra for these 12 sources, we confirmed that three of them present a $TS>25$, which is consistent with other HAWC analyses. In the case of other  two BL Lac objects with $TS>25$ (1ES 1215+303 and PG 1218+304), a more detailed analysis in needed to exclude a likely mutual contamination. \\

This work confirms the ability of HAWC to characterize the long-term VHE emission AGNs. As HAWC continues to collect gamma-ray data, we expect to improve the results on AGN long term properties in the future.

\clearpage
\section*{Full Authors List: HAWC Collaboration}
%
%
\scriptsize
\noindent
%
\vskip2cm
\noindent
A. Albert$^{1}$,
R. Alfaro$^{2}$,
C. Alvarez$^{3}$,
A. Andrés$^{4}$,
J.C. Arteaga-Velázquez$^{5}$,
D. Avila Rojas$^{2}$,
H.A. Ayala Solares$^{6}$,
R. Babu$^{7}$,
E. Belmont-Moreno$^{2}$,
K.S. Caballero-Mora$^{3}$,
T. Capistrán$^{4}$,
S. Yun-Cárcamo$^{8}$,
A. Carramiñana$^{9}$,
F. Carreón$^{4}$,
U. Cotti$^{5}$,
J. Cotzomi$^{26}$,
S. Coutiño de León$^{10}$,
E. De la Fuente$^{11}$,
D. Depaoli$^{12}$,
C. de León$^{5}$,
R. Diaz Hernandez$^{9}$,
J.C. Díaz-Vélez$^{11}$,
B.L. Dingus$^{1}$,
M. Durocher$^{1}$,
M.A. DuVernois$^{10}$,
K. Engel$^{8}$,
C. Espinoza$^{2}$,
K.L. Fan$^{8}$,
K. Fang$^{10}$,
N.I. Fraija$^{4}$,
J.A. García-González$^{13}$,
F. Garfias$^{4}$,
H. Goksu$^{12}$,
M.M. González$^{4}$,
J.A. Goodman$^{8}$,
S. Groetsch$^{7}$,
J.P. Harding$^{1}$,
S. Hernandez$^{2}$,
I. Herzog$^{14}$,
J. Hinton$^{12}$,
D. Huang$^{7}$,
F. Hueyotl-Zahuantitla$^{3}$,
P. Hüntemeyer$^{7}$,
A. Iriarte$^{4}$,
V. Joshi$^{28}$,
S. Kaufmann$^{15}$,
D. Kieda$^{16}$,
A. Lara$^{17}$,
J. Lee$^{18}$,
W.H. Lee$^{4}$,
H. León Vargas$^{2}$,
J. Linnemann$^{14}$,
A.L. Longinotti$^{4}$,
G. Luis-Raya$^{15}$,
K. Malone$^{19}$,
J. Martínez-Castro$^{20}$,
J.A.J. Matthews$^{21}$,
P. Miranda-Romagnoli$^{22}$,
J. Montes$^{4}$,
J.A. Morales-Soto$^{5}$,
M. Mostafá$^{6}$,
L. Nellen$^{23}$,
M.U. Nisa$^{14}$,
R. Noriega-Papaqui$^{22}$,
L. Olivera-Nieto$^{12}$,
N. Omodei$^{24}$,
Y. Pérez Araujo$^{4}$,
E.G. Pérez-Pérez$^{15}$,
A. Pratts$^{2}$,
C.D. Rho$^{25}$,
D. Rosa-Gonzalez$^{9}$,
E. Ruiz-Velasco$^{12}$,
H. Salazar$^{26}$,
D. Salazar-Gallegos$^{14}$,
A. Sandoval$^{2}$,
M. Schneider$^{8}$,
G. Schwefer$^{12}$,
J. Serna-Franco$^{2}$,
A.J. Smith$^{8}$,
Y. Son$^{18}$,
R.W. Springer$^{16}$,
O.~Tibolla$^{15}$,
K. Tollefson$^{14}$,
I. Torres$^{9}$,
R. Torres-Escobedo$^{27}$,
R. Turner$^{7}$,
F. Ureña-Mena$^{9}$,
E. Varela$^{26}$,
L. Villaseñor$^{26}$,
X. Wang$^{7}$,
I.J. Watson$^{18}$,
F. Werner$^{12}$,
K.~Whitaker$^{6}$,
E. Willox$^{8}$,
H. Wu$^{10}$,
H. Zhou$^{27}$

\vskip2cm
\noindent
$^{1}$Physics Division, Los Alamos National Laboratory, Los Alamos, NM, USA,
$^{2}$Instituto de Física, Universidad Nacional Autónoma de México, Ciudad de México, México,
$^{3}$Universidad Autónoma de Chiapas, Tuxtla Gutiérrez, Chiapas, México,
$^{4}$Instituto de Astronomía, Universidad Nacional Autónoma de México, Ciudad de México, México,
$^{5}$Instituto de Física y Matemáticas, Universidad Michoacana de San Nicolás de Hidalgo, Morelia, Michoacán, México,
$^{6}$Department of Physics, Pennsylvania State University, University Park, PA, USA,
$^{7}$Department of Physics, Michigan Technological University, Houghton, MI, USA,
$^{8}$Department of Physics, University of Maryland, College Park, MD, USA,
$^{9}$Instituto Nacional de Astrofísica, Óptica y Electrónica, Tonantzintla, Puebla, México,
$^{10}$Department of Physics, University of Wisconsin-Madison, Madison, WI, USA,
$^{11}$CUCEI, CUCEA, Universidad de Guadalajara, Guadalajara, Jalisco, México,
$^{12}$Max-Planck Institute for Nuclear Physics, Heidelberg, Germany,
$^{13}$Tecnologico de Monterrey, Escuela de Ingeniería y Ciencias, Ave. Eugenio Garza Sada 2501, Monterrey, N.L., 64849, México,
$^{14}$Department of Physics and Astronomy, Michigan State University, East Lansing, MI, USA,
$^{15}$Universidad Politécnica de Pachuca, Pachuca, Hgo, México,
$^{16}$Department of Physics and Astronomy, University of Utah, Salt Lake City, UT, USA,
$^{17}$Instituto de Geofísica, Universidad Nacional Autónoma de México, Ciudad de México, México,
$^{18}$University of Seoul, Seoul, Rep. of Korea,
$^{19}$Space Science and Applications Group, Los Alamos National Laboratory, Los Alamos, NM USA
$^{20}$Centro de Investigación en Computación, Instituto Politécnico Nacional, Ciudad de México, México,
$^{21}$Department of Physics and Astronomy, University of New Mexico, Albuquerque, NM, USA,
$^{22}$Universidad Autónoma del Estado de Hidalgo, Pachuca, Hgo., México,
$^{23}$Instituto de Ciencias Nucleares, Universidad Nacional Autónoma de México, Ciudad de México, México,
$^{24}$Stanford University, Stanford, CA, USA,
$^{25}$Department of Physics, Sungkyunkwan University, Suwon, South Korea,
$^{26}$Facultad de Ciencias Físico Matemáticas, Benemérita Universidad Autónoma de Puebla, Puebla, México,
$^{27}$Tsung-Dao Lee Institute and School of Physics and Astronomy, Shanghai Jiao Tong University, Shanghai, China,
$^{28}$Erlangen Centre for Astroparticle Physics, Friedrich Alexander Universität, Erlangen, BY, Germany



\end{document}